\documentclass[twocolumn,showpacs,preprintnumbers,amsmath,amssymb]{revtex4-1}

\usepackage{graphicx}
\usepackage{dcolumn}
\usepackage{bm}

\begin{document}

\preprint{APS/123-QED}

\title{Spin-wave stiffness and micromagnetic exchange interactions \\
expressed  by means of the KKR Green function approach}

\author{S.~Mankovsky, S.~Polesya  and H.~Ebert}
\affiliation{%
Department of Chemistry/Phys. Chemistry, LMU Munich,
Butenandtstrasse 11, D-81377 Munich, Germany \\
}%

\date{\today}

\begin{abstract}
We represent an approach to calculate 
micromagnetic model parameters such as the tensor of exchange
stiffness, Dzyaloshinskii-Moriya interaction as well as spin-wave
stiffness. The scheme is based on the fully relativistic 
Korringa-Kohn-Rostoker Green function 
(KKR-GF) technique
and can be seen as a relativistic extension of the
work of Lichtenstein {\em et al.} 
The expression for $D^{z\alpha}$ elements of DM differ from the
expressions for $D^{x\alpha}$ and  $D^{y\alpha}$ elements as the former are
derived via second-order perturbation term of the energy caused by
spin-spiral while the latter are associated with the first-order term. 
Corresponding numerical results are compared with
those obtained using other schemes reported in the literature. 

\end{abstract}

\pacs{71.15.-m,71.55.Ak, 75.30.Ds}
\maketitle

\section{ Introduction \label{IN}}

In order to map the DFT total energy onto the Heisenberg model and its extensions,  
different schemes have been reported in the literature
\cite{LKAG87,AKL97,Kub09,USPW03,EM09a}, giving access to 
 first principles calculations of the exchange coupling
parameters. An expression to calculate within multiple scattering
theory (MST) or, equivalently,
Korringa-Kohn-Rostoker Green function (KKR-GF) formalism
the  the isotropic exchange parameters
entering the classical Heisenberg Hamiltonian has been derived first by 
Lichtenstein et al. \cite{LKAG87}. The classical Heisenberg model has been
extended to account for relativistic effects on the inter-atomic
exchange interactions, accounting  first of all for the  Dzyaloshinskii-Moriya
interaction (DMI). An approach to calculate the
corresponding  interaction parameters,
also based on the MST formalism, was suggested by Udvardi
et al. \cite{USPW03}. Both approaches mentioned above use the magnetic force
theorem that allows to evaluate the energy change 
 associated with a distortion of the magnetisation of a system
via the expression: 
\begin{eqnarray}
\Delta {\cal E} & \approx & \int^{E_F} dE \, (E - E_F) \, \delta n(E) \;,
\label{Eq_Free_energy-1}
\end{eqnarray}
where $n(E)$ is the density of states (DOS) of the electrons.
The use of multiple scattering formalism allows the direct calculation of the  
interatomic exchange interaction using Lloyd's formula \cite{Zel08}
that gives the energy integrated DOS (NOS) $N(E)$.
This leads to an explicit expression for the energy change due to the
 tilting
of two magnetic moments in the FM ordered system and as a result to the
exchange coupling parameters. Using the extended
Heisenberg Hamiltonian  $ {\cal H}_{H}$ 
in the form suggested by Udvardi et al. \cite{USPW03} 
\begin{eqnarray}
 {\cal H}_{H} &=& -\frac{1}{2} \sum_{i \neq j}
 \mathbf{e}_i{\underline{J}}_{ij}\mathbf{e}_j  + \sum_{i}
 K(\mathbf{e}_i) 
\label{Eq_Heisenberg}
\end{eqnarray}
the isotropic exchange interaction and DMI
parameters  are deduced from the  symmetric
and antisymmetric parts of the exchange tensor $\underline{J}_{ij}$:
\begin{eqnarray*}
J_{ij} &=& \frac{1}{3} \mbox{Tr}  \, \underline{J}_{ij}
\end{eqnarray*}
and
\begin{eqnarray*}
D^{\nu}_{ij} &=&  \epsilon^{\lambda\mu\nu}
 \frac{J^{\lambda\mu}_{ij} - J^{\mu\lambda}_{ij}}{2}  \; ,
\end{eqnarray*} 
with $\epsilon^{\lambda\mu\nu}$ the Levi-Civita tensor.
A similar formulation for the exchange tensor, also on the basis of MST, has been
suggested in our previous work \cite{EM09a}.

Adopting a micromagnetic approach the free energy density 
may be expressed by \cite{BH94}
\begin{eqnarray}
  F(\vec{r}) &=& \sum_{\alpha}
                 A_{\alpha\alpha}\left(\frac{\partial \hat{m}
                 }{\partial r_{\alpha}}\right)^2 + 
                 \sum_{\alpha\nu}
                 {D}^{\nu \alpha} 
  \bigg(\hat{m} \times  \frac{\partial \hat{m}}{\partial r_\alpha} \bigg)_{\nu} \;.           
\label{Eq_Micromag}
\end{eqnarray}
Also in this case, the various parameters can be evaluated from first principles
calculations. In particular, the spin-wave stiffness ${\mathfrak D}_{\alpha\alpha}$
can be evaluated from the second order derivative of the spin-spiral energy
$E (\vec{q})$\cite{Kueb00}:  
\begin{eqnarray} 
 \mathfrak{D}_{\alpha\alpha}  &=&  
\frac{2g}{M}  \frac{\partial^2 E(\vec{q}) }{\partial q_\alpha \partial
  q_\alpha}  \;.
\label{Eq_deriv_dqdq}
\end{eqnarray}
The corresponding expression for the  closely connected 
 exchange stiffness $A_{\alpha\alpha} =
{\mathfrak D}_{\alpha\alpha}M/(2g)$ (where $g$ is the Land\'{e} factor
and $M$ is the total magnetic moment  \cite{HGM+09}) 
has been derived by
Liechtenstein et al.\cite{LKG84} by means of
 non-relativistic multiple scattering theory. 

Recently,  Freimuth et al. \cite{FBM14}  demonstrated that
the parameters entering the relativistic free energy density in Eq.\ 
(\ref{Eq_Micromag}), i.e.\ the Dzyaloshinskii-Moriya interaction and 
exchange stiffness, can be computed by using the Berry phase approach.
The microscopic 
 DMI parameters in this case are evaluated as the slope 
of the spin-wave energy $E(\vec{q})$ at $\vec{q} = 0$:
\begin{equation}
D^{\nu\alpha}  = \left(\frac{\partial E\big((\hat{z}\times\delta\hat{m}(\vec{q}))_\nu\big)}{\partial q_\alpha}\right)_{q = 0} \; .
\label{Eq_deriv_dq}
\end{equation}

In the present work we represent an approach for the calculation of the
parameters of the Heisenberg and micromagnetic models performed on the same footing
within the fully-relativistic spin-polarized Korringa-Kohn-Rostoker
Green function (KKR-GF) method.

\section{Theoretical background based on the spin spiral
  approach \label{TB}}

\subsection{Representation of the electronic structure}

To derive explicite expressions for the various interaction parameters
on the basis of electronic structure calculations, we start from 
the Dirac Hamiltonian set up within the framework 
of relativistic spin-density functional theory  \cite{MV79}:
\begin{eqnarray}
{\cal H}_{\rm D}  & =&
- i c \vec{\bm \alpha} \cdot \vec \nabla  + \frac{1}{2} \, c^{2} ({\bm\beta} - 1) + V(\vec r)  + \beta \vec{\bm\sigma}\cdot {\vec B}_{xc}(\vec r)
\; ,
\label{Hamiltonian}           
\end{eqnarray}
%
where ${\vec B}_{xc}(\vec r)$ is the spin-dependent part of the 
exchange-correlation potential and all other quantities have there
usual meaning \cite{Ros61,EBKM16}.

Instead of representing the electronic structure in terms 
of Bloch states derived from the Hamiltonian 
in Eq.\ (\ref{Hamiltonian}) it is much more convienient for our 
purposes to use the electronic Green function 
$G(\vec{r},\vec{r}\,',E)$ instead.
Within the KKR-GF approach $G(\vec{r},\vec{r}\,',E)$  
is represented in real space by the expression \cite{EBKM16}:
\begin{eqnarray}
G(\vec{r},\vec{r}\,',E) & = &
\sum_{\Lambda_1\Lambda_2} 
Z^{n}_{\Lambda_1}(\vec{r},E)
                              {\tau}^{n n'}_{\Lambda_1\Lambda_2}(E)
Z^{n' \times}_{\Lambda_2}(\vec{r}\,',E)
 \nonumber \\
 & & 
-  \sum_{\Lambda_1} \Big[ 
Z^{n}_{\Lambda_1}(\vec{r},E) J^{n \times}_{\Lambda_1}(\vec{r}\,',E)
\Theta(r'-r)  \nonumber 
\\
 & & \qquad\quad 
J^{n}_{\Lambda_1}(\vec{r},E) Z^{n \times}_{\Lambda_1}(\vec{r}\,',E) \Theta(r-r')
\Big] \delta_{nn'} \; .
\label{Eq_KKR-GF}
\end{eqnarray}
Here 
$Z^{n}_{\Lambda_1}(\vec{r},E)$ and 
$J^{n}_{\Lambda_1}(\vec{r},E)$ 
are
the regular and irregular  solutions of the
single site Dirac equation and
  ${\underline{\tau}}^{n n'}$ is the 
  so-called scattering path operator matrix \cite{EBKM16}.

The specific form of the Dirac Hamiltonian 
in Eq.\ (\ref{Hamiltonian}) also allows to express the 
impact of the change in the potential $\Delta V(\vec{r})$ 
due to the rotation 
of the magnetic moments on the atomic sites in a very simple way.
Assuming that  ${\vec B}_{xc}(\vec r)$ 
on site $i$ is aligned 
along the orientation of the spin moment $\hat{m}_i$, 
i.e.\  $\vec{B}_{xc}(\vec{r}) = {B}_{xc}(\vec{r}) \hat{m}_i$,
 and taking into account that $ \hat{m}_i = \hat{z}$ 
 for a ferromagnetic (FM)  state,  the 
potential change $\Delta V(\vec{r})$ 
connected with the tilting of rigid magnetic moments 
has the form
\begin{eqnarray}
 \Delta V(\vec{r}) &=&  \sum_i \beta \big( \vec{\sigma}\cdot\hat{m}_i
  -  \sigma_z\big) B_{xc}(\vec{r}) \;.
\label{Eq_perturb_stiff}
\end{eqnarray}

\subsection{Basic properties of the exchange interactions}

Similarly to our previous work \cite{EM09a}, the present approach is based
on the magnetic force theorem. As a starting point we use the ferromagnetic (FM) state as a reference
state and neglect for the moment all temperature effects, 
i.e.\ assume $T = 0$ K.
In this case, a change of the grandcanonical potential caused by the
formation of a spin spiral in the system is given by Eq.\ (\ref{Eq_Free_energy-1}).
However, instead of using the Lloyd formula, we represent the change
of the density of states $\Delta n(E)$ in terms of the Green function
$G_0(E)$ for the FM reference state, which is modified due to the perturbation.
Denoting the corresponding change in the Green function $\Delta G(E)$
one can write: 
\begin{eqnarray}
\Delta {\cal E} & \approx &  -\frac{1}{\pi} 
\mbox{Im}\, \mbox{Tr}  \int^{E_F} dE \,
(E - E_F) \, \Delta G(E) \; .
\label{Eq_DeltaF_1}
\end{eqnarray}

Assuming that the perturbation is small, the induced 
change of the Green function can be represented 
by the following perturbation expansion
\begin{eqnarray}
\Delta G(E) & = & G_0(E) V G_0(E) +  G_0(E) V G_0(E) V G_0(E) + ... \;,
\label{Eq_GF_expansion}
\end{eqnarray}
where $V$  is a perturbation operator describing the spin-spiral
creation in the FM system. 
Substituting Eq.\ (\ref{Eq_GF_expansion}) into Eq.\
(\ref{Eq_DeltaF_1}) and using the sum rule $\frac{dG}{dE} = - GG$ for the
Green function, one obtains an expression for the free energy change
associated with the spin spiral:
\begin{eqnarray}
\Delta {\cal E} &=&   -\frac{1}{\pi} \mbox{Im}\,\mbox{Tr} \int^{E_F} dE\, (E - E_F) \,
G_0(E) \, V \, G_0(E) \nonumber \\ 
&&-\frac{1}{\pi} \mbox{Im}\,\mbox{Tr} \int^{E_F} dE\, V
   G_0(E)  V G_0(E) \\
   &=& K^{(1)} + K^{(2)}   \;.
\label{Eq_Free_Energy}
\end{eqnarray}
Here only the
first- and second-order terms of the expansion are kept as they are
responsible for the effects discussed below. Note however, that
higher-order terms can also be non-negligible leading to corresponding
higher-order exchange interaction terms in the Heisenberg Hamiltonian,
which however are not discussed in the present work.

\subsection{Spin-wave stiffness }

We first consider a spin spiral characterized by
its wave vector $\vec{q}$, with the magnetic moment direction on site
$(i)$ given by the expression
\begin{eqnarray}
  \hat{m}_i &=&
 (\mbox{sin} \theta \;\mbox{cos}(\vec{q} \cdot \vec{R}_i),
 \mbox{sin} \theta \;\mbox{sin}(\vec{q} \cdot
 \vec{R}_i),\mbox{cos} \theta ) \;,
\label{spiral2}
\end {eqnarray}
implying the same cone angle $\theta$ for all atomic sites.

For the sake of convenience, we start with the Heisenberg model.
When the spin-spiral is given in the form of Eq.\ (\ref{spiral2}) one gets the
following change in energy  $ \Delta E_{H}$
with respect to the FM 
state with its magnetization direction along $\hat{z}$
\begin{eqnarray}
  \Delta E_{H} &=&  - \theta \sum_{i \neq j}^N
                          D^x_{ij} [\mbox{sin}(\vec{q} \cdot \vec{R}_i)
                         - \mbox{sin}(\vec{q} \cdot \vec{R}_j) ]  \nonumber \\
  & & - \theta ^2 \sum_{ij}^N
                         [ D^z_{ij} \mbox{sin}(\vec{q} \cdot (\vec{R}_j
  -\vec{R}_i) \; \;   \;\nonumber  \\
 && + \frac{1}{2}[(J^{xy}_{ij} + J^{yx}_{ij})  \mbox{sin}(\vec{q} \cdot
  (\vec{R}_j + \vec{R}_i) \; \; \nonumber  \\    
 &&  + (J^{xx}_{ij} - J^{yy}_{ij}) \mbox{cos}(\vec{q} \cdot (\vec{R}_j
  + \vec{R}_i) \; \;   \;\nonumber  \\
 && + (J^{xx}_{ij} + J^{yy}_{ij})  \mbox{cos}(\vec{q} \cdot
  (\vec{R}_j -\vec{R}_i)] \; + ... \; ,  
\label{Eq_Heisenberg_spin-spiral}
\end{eqnarray}
where we restrict the expansion up to second order with respect to the
angle $\theta$, and focus in the following
on the term proportional to $\theta^2$.

Using the expressions for the Green function given in Eq.\
(\ref{Eq_KKR-GF}) and for the perturbation 
due to the formation of a spin spiral according to 
 Eq.\
(\ref{Eq_perturb_stiff}), one obtains an expression for the free energy
contribution $K^{(2)}$ given in the multiple scattering 
representation
\begin{widetext}  
%
\begin{eqnarray}
 K^{(2)}  &=& 
  -\frac{\theta ^2}{\pi}  \sum_{i \neq j}^N \mbox{Im} \, \mbox{Tr} \int 
dE              \nonumber \\
&&\bigg[\frac{1}{2}( \underline{T}_x^i \; \underline{\tau}^{ij}\;  \underline{T}_x^j\; \underline{\tau}^{ji}\; + \underline{T}_y^i \; \underline{\tau}^{ij}\;  \underline{T}_y^j\; \underline{\tau}^{ji}\;)  \mbox{cos}(\vec{q} \cdot (\vec{R}_j -\vec{R}_i) \;) \; \nonumber \\
&&+ \frac{1}{2}( \underline{T}_x^i \; \underline{\tau}^{ij}\;  \underline{T}_x^j\; \underline{\tau}^{ji}\; - \underline{T}_y^i \; \underline{\tau}^{ij}\;  \underline{T}_y^j\; \underline{\tau}^{ji}\;)  \mbox{cos}(\vec{q} \cdot (\vec{R}_j + \vec{R}_i) \;) \;
   \nonumber  \\
&&+ \frac{1}{2}( \underline{T}_x^i \; \underline{\tau}^{ij}\;  \underline{T}_y^j\; \underline{\tau}^{ji}\; - \underline{T}_y^i \; \underline{\tau}^{ij}\;  \underline{T}_x^j\; \underline{\tau}^{ji}\;)  \mbox{sin}(\vec{q} \cdot (\vec{R}_j -\vec{R}_i) \;) \;   \;\nonumber  \\
&&+ \frac{1}{2}( \underline{T}_x^i \; \underline{\tau}^{ij}\;  \underline{T}_y^j\; \underline{\tau}^{ji}\;+ \underline{T}_y^i \; \underline{\tau}^{ij}\;  \underline{T}_x^j\; \underline{\tau}^{ji}\;)  \mbox{sin}(\vec{q} \cdot (\vec{R}_j + \vec{R}_i)) \; \;\bigg].
\label{Denergy1}
\end{eqnarray}
Considering for the sake of simplicity a system with one atom per unit cell one has
the following matrix elements
representing the change in the  potential
\begin{eqnarray}
 T_{x,\Lambda_1\Lambda_2}(E) & = & \int_{\Omega_0} d^3r
                                   Z^{\times}_{\Lambda_1}(\vec{r}\;,E)\;
                                   \beta\, \sigma_x\, B_{xc}(\vec{r\;})
                                   Z_{\Lambda_2}(\vec{r}\;,E) \nonumber \\
 T_{y,\Lambda_1\Lambda_2}(E) & = & \int_{\Omega_0} d^3r
                                   Z^{\times}_{\Lambda_1}(\vec{r}\;,E)\;
                                   \beta\, \sigma_y\, B_{xc}(\vec{r\;})
                                   Z_{\Lambda_2}(\vec{r}\;,E)
\label{Eq:TORQUE-ME}
\; ,
\end{eqnarray}
with  $\underline{T}_{\mu}^i =   \underline{T}_{\mu}^j =
\underline{T}_{\mu} $.
By doing a one-to-one comparison of the energy terms associated with a pair of
sites $(i, j)$
as given by Eqs.\ (\ref{Eq_Heisenberg_spin-spiral}) and (\ref{Denergy1}), 
respectively, one gets expressions for the elements $J^{yx}_{ij}$ and
$J^{xy}_{ij}$ of the $\underline{J}_{ij}$ tensor, as well as the
interatomic DMI terms $D^z_{ij}$, that have the
same form as those derived previously by us using the Lloyd formula \cite{EM09a}.

\medskip

In contrast to our previous work, the goal of the present study is to get
expressions for the micromagnetic interaction parameters.
This is achieved by performing a Fourier transformation for the
scattering path operator 
\begin{eqnarray}
\underline{\tau}^{ij} &=& \frac{1}{\Omega_{BZ}} \int d^3k
e^{i\vec{k}(R_j-R_i)}\underline{\tau}(\vec{k})
\label{Eq:TAU-FT}
\end{eqnarray}
leading to an alternative expression for the energy change caused
by the formation of a spin-spiral 
\begin{eqnarray}
 K^{(2)}  &=&  -\frac{\theta ^2}{2\pi} \mbox{Im} \mbox{Tr} \int 
dE \frac{1}{\Omega_{BZ}}\int d^3k \frac{1}{\Omega_{BZ}}\int d^3k' \nonumber \\
&&\times \bigg[\frac{1}{2}( \underline{T}_x \; \underline{\tau}(\vec{k},E)\;  \underline{T}_x\; \underline{\tau}(\vec{k}',E)\; + \underline{T}_y \; \underline{\tau}(\vec{k},E)\;  \underline{T}_y\; \underline{\tau}(\vec{k}',E)\;) (\delta(\vec{k} + \vec{q} -\vec{k}') + \delta(\vec{k} - \vec{q} -\vec{k}') )   \;\nonumber\\
&&+ \frac{1}{2i}( \underline{T}_x \; \underline{\tau}(\vec{k},E)\;  \underline{T}_y\; \underline{\tau}(\vec{k}',E)\; - \underline{T}_y \; \underline{\tau}^(\vec{k},E)\;  \underline{T}_x\; \underline{\tau}(\vec{k}',E)\;) (\delta(\vec{k} + \vec{q} -\vec{k}') - \delta(\vec{k} - \vec{q} -\vec{k}') )  \;\nonumber\\
&&+ \frac{1}{2}( \underline{T}_x \; \underline{\tau}(\vec{k},E)\;  \underline{T}_x\; \underline{\tau}(\vec{k}',E)\; - \underline{T}_y \; \underline{\tau}(\vec{k},E)\;  \underline{T}_y\; \underline{\tau}(\vec{k}',E)\;) (\delta(\vec{k} + \vec{q} -\vec{k}') + \delta(\vec{k} - \vec{q} -\vec{k}') )\delta(\vec{k} - \vec{k}') \; \nonumber\\
&&+ \frac{1}{2i}( \underline{T}_x \; \underline{\tau}(\vec{k},E)\;  \underline{T}_y\; \underline{\tau}(\vec{k}',E)\;+ \underline{T}_y \; \underline{\tau}(\vec{k},E)\;  \underline{T}_x\; \underline{\tau}(\vec{k}',E)\;) (\delta(\vec{k} + \vec{q} -\vec{k}') - \delta(\vec{k} - \vec{q} -\vec{k}') )\delta(\vec{k} - \vec{k}') \;
 \; \bigg] 
 \;.
\label{Denergy3}
\end{eqnarray}

The last term in Eq.\ (\ref{Denergy3}) is equal to zero, while the third one
corresponds to the non-local MCA discussed by Udvardi et al. \cite{USPW03},
which gives for the present geometry the contribution to the MCA within
the $xy$-plane
 \begin{eqnarray}
 \Delta E_{MCA}  &=&   -\frac{\theta^2}{2\pi} \mbox{Im} \, \mbox{Tr} \int 
dE \frac{1}{\Omega_{BZ}}\int d^3k  \nonumber \\
&&\times [ \underline{T}_x \; \underline{\tau}(\vec{k},E)\;  \underline{T}_x\; \underline{\tau}(\vec{k},E) - \underline{T}_y \; \underline{\tau}(\vec{k},E)\;  \underline{T}_y\;\underline{\tau}(\vec{k},E)    \;  ] \;
\label{Denergy2}
\; .
\end{eqnarray}

To consider the first two terms in Eq.\ (\ref{Denergy3}), one can use
a Taylor expansion of the $\tau$ matrix for small $\vec{q}$-vectors

\begin{eqnarray}
\underline{\tau}(\vec{k} \pm \vec{q},E) &=& \underline{\tau}(\vec{k},E) \pm \sum_{\alpha} \frac{\partial \underline{\tau}(\vec{k},E) }{\partial k_\alpha}q_{\alpha}  + \frac{1}{2} \sum_{\alpha,\beta} \frac{\partial^2 \underline{\tau}(\vec{k},E) }{\partial k_\alpha \partial k_\beta}q_{\alpha}q_{\beta}
\;,
\end{eqnarray}
that gives the corresponding contribution  $K^{(2)}_{1-2}$ to $ K^{(2)}$
\begin{eqnarray}
  K^{(2)}_{1-2} &=&  -\frac{\theta ^2}{2\pi} \mbox{Im}\; \mbox{Tr} \int 
dE \frac{1}{\Omega_{BZ}}\int d^3k  \nonumber \\
&&\frac{1}{2}\Bigg[\underline{T}_x \; \underline{\tau}(\vec{k},E)\;
   \underline{T}_x\; \Bigg(2 \underline{\tau}(\vec{k},E) +
   \underbrace{\sum_{\alpha,\beta} \frac{\partial^2 \underline{\tau}(\vec{k},E)
   }{\partial k_\alpha \partial k_\beta}q_{\alpha}q_{\beta} }_{T1} \Bigg)\;
+ \underline{T}_y \; \underline{\tau}(\vec{k},E)\;  \underline{T}_y\;
\Bigg(2 \underline{\tau}(\vec{k},E) +  \underbrace{ \sum_{\alpha,\beta} \frac{\partial^2
   \underline{\tau}(\vec{k},E) }{\partial k_\alpha \partial
  k_\beta}q_{\alpha}q_{\beta}}_{T2} \Bigg)\; \nonumber
\\
&&+ \frac{1}{i}\Bigg( \underline{T}_x \; \underline{\tau}(\vec{k},E)\;
\underline{T}_y\; \sum_{\alpha} 2\; \frac{\partial
  \underline{\tau}(\vec{k},E) }{\partial k_\alpha}q_{\alpha}  \; -
\underline{T}_y \; \underline{\tau}(\vec{k},E)\;
\underline{T}_x\; \sum_{\alpha} 2\; \frac{\partial
  \underline{\tau}(\vec{k},E) }{\partial k_\alpha}q_{\alpha} \Bigg) \Bigg] \;.
\label{Denergy_K2}
\end{eqnarray}
Doing an integration by parts for the expression involving the term $T1$ 
indicated in Eq.\
(\ref{Denergy_K2}) (see \cite{MWE18}), one obtains 
\begin{eqnarray}
  T1  &=&  -\frac{\theta ^2}{2\pi} \mbox{Im} \, \mbox{Tr} \int 
dE \frac{1}{\Omega_{BZ}}\int d^3k \frac{1}{2} \underline{T}_x \; \underline{\tau}(\vec{k},E)\;
\underline{T}_x\;  \sum_{\alpha,\beta} \frac{\partial^2
  \underline{\tau}(\vec{k},E) }{\partial k_\alpha \partial
  k_\beta}q_{\alpha}q_{\beta} \; \nonumber  \\
&=&  \frac{\theta ^2}{4\pi} \sum_{\alpha,\beta}
q_{\alpha}q_{\beta} \mbox{Im}  \, \mbox{Tr} \int 
dE \frac{1}{\Omega_{BZ}}  \int d^3k \underline{T}_x \; \frac{\partial
  \underline{\tau} (\vec{k},E)}{\partial k_\alpha} \;
\underline{T}_x\; \frac{\partial \underline{\tau}(\vec{k},E) }{\partial
  k_\beta}\;.
\label{Denergy1-X}
\end{eqnarray}
The same transformation can also be made for the term $T2$.
\medskip

Equating now 
the second order derivatives with respect to the $\vec{q}$ vector for the
microscopic and model energies in the limit ${q} \to 0$
\begin{equation}
\left(\frac{\partial^2 \Delta {\cal E}}{\partial q_\alpha \partial q_\beta}\right)_{q = 0} 
=\left(\frac{\partial^2 { K^{(2)}}}{\partial q_\alpha \partial q_\beta}\right)_{q = 0} 
= \left(\frac{\partial^2 \Delta E_{H}}{\partial q_\alpha \partial q_\beta}\right)_{q = 0}
\label{Eq_qq_derivatives}
\end{equation}
one  obtains the components of the exchange tensor.
This leads to an expression for the spin-wave stiffness 
\begin{eqnarray}
 {\mathfrak D}_{\alpha\beta}  &=&    \frac{1}{\theta^2} \frac{4}{M}  \frac{\partial^2 E }{\partial q_\alpha \partial
  q_\beta}  \nonumber  \\
&=& \frac{1}{\pi M} \mbox{Im}\;
 \mbox{Tr} \int dE \;\; \frac{1}{\Omega_{BZ}}\int d^3k \left[ \underline{T}_x \;\frac{\partial \underline{\tau}(\vec{k},E)}{\partial
  k_\alpha}\;  \underline{T}_x\; \frac{\partial
   \underline{\tau}(\vec{k},E)}{\partial k_\beta}\; + \underline{T}_y \; \;\frac{\partial \underline{\tau}(\vec{k},E)}{\partial
  k_\alpha}\;  \underline{T}_y\;  \frac{\partial
   \underline{\tau}(\vec{k},E)}{\partial k_\beta}\;\right] \, ,
\label{Eq_sw-stiffness}
\end{eqnarray}
that can be seen as a relativistic generalization of the expression given by 
Liechtenstein et al.\ \cite{LKAG87}.

\subsection{Dzyaloshinskii-Moriya interaction}

Taking the first-order derivative of $ K^{(2)}_{1-2}$  with respect to the components of the $\vec{q}$-vector
 the last term in Eq.\ (\ref{Denergy3}) 
  gives in the limit $q \to 0$ the elements $D^{z\alpha}$ of the DMI tensor:
\begin{eqnarray}
D^{z\alpha} &=& \frac{1}{\theta^2}   \lim_{q \to 0} \frac{\partial
  K^{2}}{\partial q_\alpha} \nonumber \\
 &=&  \left(\frac{1}{2\pi} \right)\mbox{Re}\; \mbox{Tr} \int 
dE \frac{1}{\Omega_{BZ}}\int d^3k \left[ \underline{T}_x \; \underline{\tau}(\vec{k},E)\;  \underline{T}_y\; \frac{\partial \underline{\tau}(\vec{k},E) }{\partial k_\alpha}  \; - \underline{T}_y \; \underline{\tau}(\vec{k},E)\;  \underline{T}_x\; \frac{\partial \underline{\tau}(\vec{k},E) }{\partial k_\alpha}  \right] \;.
\label{Eq_MM_DMI_zz}
\end{eqnarray}

\end{widetext}
In order to calculate the tensor elements $D^{x\alpha}$ and $D^{y\alpha}$,
it is convenient to use a spin spiral given in the following rather
general form  \cite{ME17}: 
\begin{equation}
  \hat{m}_i =  \hat{m}_\mu \mbox{sin}(\vec{q}\cdot\vec{R}_i) +  \hat{m}_z \mbox{cos}(\vec{q}\cdot\vec{R}_i) \; ,
\label{spiral_DM_XY}
\end {equation}
where $ \hat{m}_i$ characterizes the direction of the magnetic moments
on site $R_i$ 
with $ \hat{m}_i \equiv \hat{m}(\vec{R}_i$), $\mu = \{x, y\}$
and the wave vector $\vec{q}$ can have any direction.
Assuming a weak deviation of the magnetic
moments $\vec{m}_i$ from the $\hat{z}$ direction, 
this setting allows to get rid of the first order
derivatives with respect to $q_\alpha$,
 related to the term $K^{(2)}$  in Eq.\
(\ref{Eq_Free_Energy}) and to focus on the term $K^{(1)}$.

With this, the elements  $D^{\mu\alpha}$ of the micromagnetic tensor 
 representing the  DMI
 as defined by  Eq.\ (\ref{Eq_deriv_dq})
 are determined exclusively  by the first-order term $K^{(1)}$ in
Eq.\ (\ref{Eq_Free_Energy}). The term  $K^{(1)}$ 
associated with the perturbation
Eq.\ (\ref{Eq_perturb_stiff}) induced by a spin spiral
as described by  Eq.\  (\ref{spiral_DM_XY}) has the following form: 
\begin{widetext}
\begin{eqnarray}
  K^{(1)} & = &-  \frac{1}{\pi} \sum_{i \neq j}\mbox{Im}\, \mbox{Tr}
  \int^\mu dE (E - \mu)  \nonumber  \\
    &&\times  \Bigg( \sin(\vec{q} \cdot (\vec{R}_i - \vec{R}_j)) \Big[\underbrace{ \underline{O}^{j}(E)\,
\underline{\tau}^{ji}(E)  \underline{T}^{i}_\mu(E)\, \underline{\tau}^{i
   j}(E) }_{T1} - \underbrace{ \underline{T}^{j}_\mu(E)\,
\underline{\tau}^{ji}(E)  \underline{O}^{i}(E)\, \underline{\tau}^{i
   j}(E) }_{T2} \Big]\nonumber \\
 &&+ [ \cos(\vec{q}
\cdot (\vec{R}_i - \vec{R}_j)) -1] \Big[\underbrace{  \underline{O}^{j}(E)\, \underline{\tau}^{ji}(E)\,
\underline{T}^{i}_z(E)\, \underline{\tau}^{ij}(E)  }_{T3} - \underbrace{  \underline{T}^{j}_z(E)\, \underline{\tau}^{ji}(E)\,
\underline{O}^{i}(E)\, \underline{\tau}^{ij}(E) }_{T4} \Big]   \Bigg) \;.
\label{Eq_Delta_K_1_KKR}
\end{eqnarray}
In the case of one atom per unit cell one has
for the matrices occuring in Eq.\ (\ref{Eq_Delta_K_1_KKR})
$\underline{O}^{i}(E) = \underline{O}(E)$ and 
$\underline{T}^{i}_\mu(E) = \underline{T}_\mu(E)$,
with  $\underline{T}_\mu(E)$  given by Eq.\ (\ref{Eq:TORQUE-ME}) and
the overlap matrix given by:
\begin{eqnarray}
 O_{\Lambda_1\Lambda_2}(E) & = & \int_{\Omega_0} d^3r
 Z^{\times}_{\Lambda_1}(\vec{r},E) Z_{\Lambda_2}(\vec{r},E)
 \; .
\label{Eq:ME}
\end{eqnarray}
Calculating the  derivative  $\frac {\partial K^{(1)}}{\partial q_y}$
in the limit ${q \to 0}$, the terms $T1$ and $T2$ in Eq.\
(\ref{Eq_Delta_K_1_KKR}) giving the only non-vanishing 
contributions to the DMI parameters are:
%
\begin{eqnarray}
T1 & \to &  -\frac{1}{\pi} \,  \lim_{q \to 0} \frac {\partial}{\partial q_\alpha}
  \Big[\mbox{Im}\, \mbox{Tr}\,
   \frac{1}{2i}  \int^\mu dE \,(E - \mu)  \nonumber \\
  & & \times \Big\{ \underline{O}(E)\, 
 \frac {1}{\Omega_{BZ}}\int d^3k \,
\underline{\tau}(\vec{k},E)\,  \underline{T}_{\mu}(E)\,
\underline{\tau}(\vec{k} - \vec{q},E) - \underline{O}(E)\, \frac {1}{\Omega_{BZ}} \int d^3k \,
\underline{\tau}(\vec{k},E)\, \underline{T}_{\mu}(E)\,
 \underline{\tau}(\vec{k} + \vec{q},E)\Big\} \Big] 
\\
T2 & \to &  -\frac{1}{\pi}  \,  \lim_{q \to 0} \frac {\partial}{\partial q_\alpha}
  \Big[\mbox{Im}\, \mbox{Tr}\,
   \frac{1}{2i}  \int^\mu dE \,(E - \mu)  \nonumber \\
  & & \times \Big\{ \underline{T}_{\mu}(E)\,  
 \frac {1}{\Omega_{BZ}}\int d^3k \,
\underline{\tau}(\vec{k},E)\, \underline{O}(E)\,
\underline{\tau}(\vec{k} - \vec{q},E) - \underline{T}_{\mu}(E)\, \frac {1}{\Omega_{BZ}} \int d^3k \,
\underline{\tau}(\vec{k},E)\, \underline{O}(E) \,
 \underline{\tau}(\vec{k} + \vec{q},E)\Big\} \Big] \;.
\end{eqnarray}
%
%
Equating  for  the microscopic and model energies the derivatives 
 with
respect to components of the $\vec{q}$-vector 
one obtains 
in the limit $\vec{q} \to 0$ 
the  elements $D^{\mu\alpha}$ of the micromagnetic DMI tensor:
\begin{eqnarray}
 D^{\mu\alpha}  &=&  \lim_{q \to 0} \frac {\partial}{\partial q_\alpha} K^{(1)} =
 \epsilon_{\mu\nu}\frac{1}{\pi} \mbox{Re}\, \mbox{Tr}\, \int^\mu dE \,(E - \mu)  \nonumber \\ 
 &&\times  \frac {1}{\Omega_{BZ}} \int d^3k \, \Big[ \underline{O}(E)\,
\underline{\tau}(\vec{k},E) \,
\underline{T}_{\nu}(E)\,\frac{\partial}{\partial k_\alpha}\,
\underline{\tau}(\vec{k},E) -  \underline{T}_{\nu}(E) \,
\underline{\tau}(\vec{k},E) \,
\underline{O}(E)\,\frac{\partial}{\partial k_\alpha}\,
\underline{\tau}(\vec{k},E)  \Big] \;
\label{Eq:DMI_XY} 
\end{eqnarray}
\end{widetext}
with $\mu =  \{x,y\}$ and $\nu = \{x,y\}$  and
$\epsilon_{\mu\nu}$ the elements of the transverse Levi-Civita tensor
 $ \underline{\epsilon}  = \begin{bmatrix}
   0 & 1 \\
   -1 & 0     
 \end{bmatrix} $.

This formulation obviously  gives access to a discussion
of the DMI parameters
 in terms of specific features of the electronic band structure in a similar
way as suggested in Ref.\ \cite{KNA15}. As 
the present formulation  is given within the KKR-GF
formalism, it allows to deal both with ordered and disordered 
materials, where disorder may be treated using the coherent potential
approximation (CPA) alloy theory.

In addition, it is worth noting that only the 
 elements $D^{x\alpha}$ and $D^{y\alpha}$
 are defined for the FM state with its magnetization
along $\hat{z}$ direction, as only the $x$- and $y$-directions are  allowed
for a change of the   transverse spin moment component. 
The elements $D^{x\alpha}$ and $D^{y\alpha}$ originate 
 from the interatomic DMI components $D^{x}_{ij}$ and $D^{y}_{ij}$, respectively,  
characterizing for $\hat{m} || \hat{z}$ the 
non-zero magnetic torques acting on one atom $i(j)$ from 
another atom at site $j(i)$ (see, e.g., Ref. \cite{MBM+09}).
As it follows from Eq.\ (\ref{Eq_Heisenberg_spin-spiral}), the terms
related to $D^{x}_{ij}$ and $D^{y}_{ij}$ appear 
in first-order within an 
 expansion of the energy with respect to the angle 
$\theta$ characterizing the deviation of a magnetic moment 
from the $\hat{z}$ direction.
On the other hand, the DMI component $D^{z}_{ij}$ 
and analogously the element $D^{zz}$ of the DMI tensor 
give the  contribution to the energy
which is of the order of $\theta^2$: 
since the DMI component $D^{z}_{ij}$ couples the 
components of the magnetic moments of atoms $i$ and $j$,
$m^x_{i,j} \sim \theta$ and 
$m^y_{i,j} \sim \theta$, respectively, 
 both components should be non-zero to give a contribution to
the energy change;
 in contrast to the terms  $D^{x(y)}_{ij}$ that couple
the   componentss $m^z$ and $m^{x(y)}_{i,j} \sim \theta$ of the magnetic moments.

\section{Results \label{res}}

\begin{figure}[h]
\includegraphics[width=0.42\textwidth,angle=0,clip]{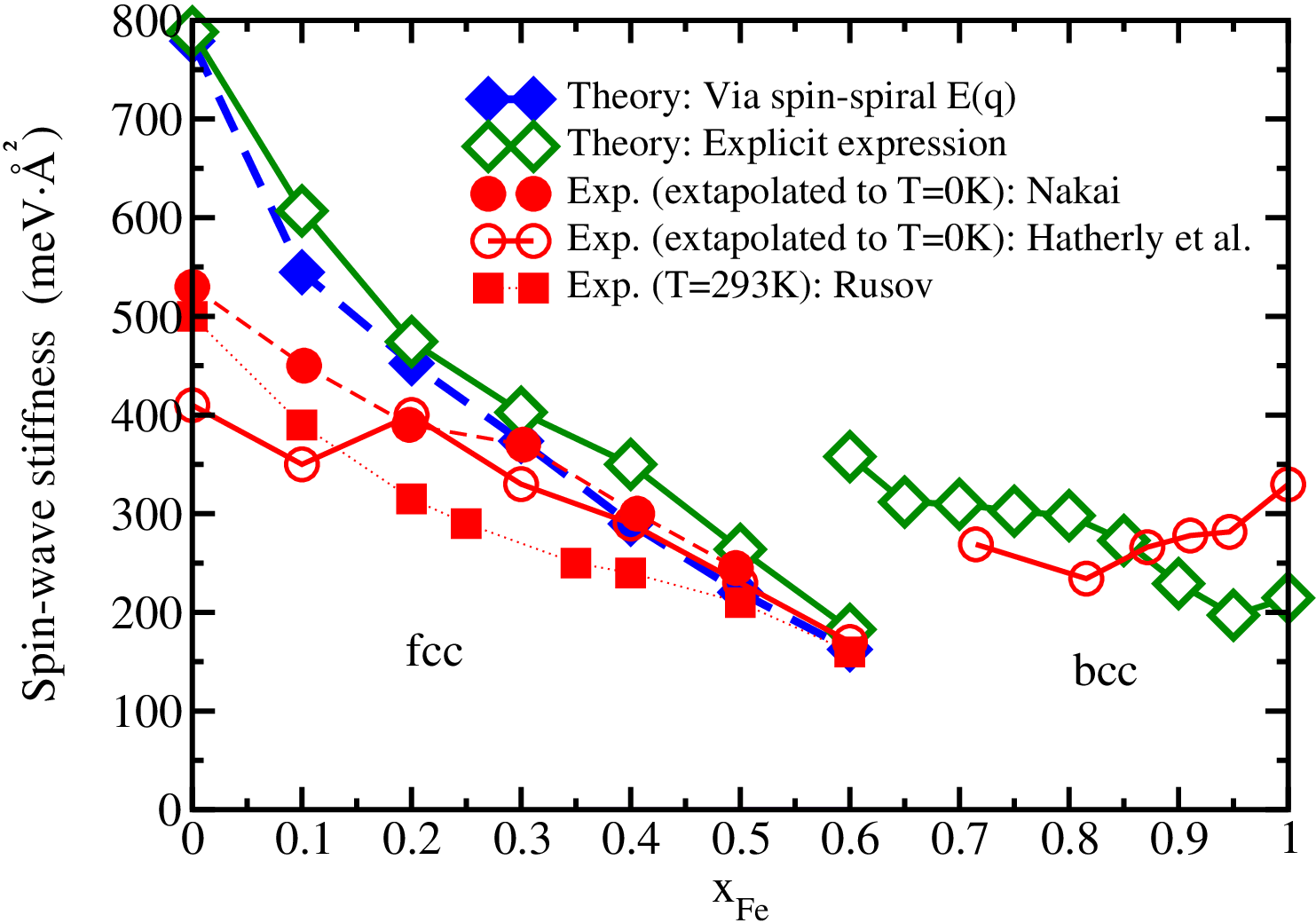}\;
\caption{\label{fig:FeNi_stiff}
Calculated spin-wave stiffness $\mathfrak{D}_{xx}$ (see text)
 of  Fe$_{1-x}$Ni$_x$ alloys, with fcc (left) and bcc
(right) crystal structures, represented as a function of the
concentration in comparison with experiment: Nakai \cite{Nak83} (filled circles),
Hatherly et al. \cite{HHL+64} (open circles), and Rusov \cite{Rus76} (filled squares). }  
\end{figure}

Figure \ref{fig:FeNi_stiff} represents 
the spin-wave stiffness parameter $\mathfrak{D}_{xx}$
for the bcc and fcc phases of disordered Fe$_{1-x}$Ni$_x$  alloys
 calculated using Eq.\ (\ref{Eq_sw-stiffness}) (open diamonds). 
%
%
In the case of the fcc-alloys, the results are compared with
the spin-wave stiffness deduced from the energy dispersion $E(\vec{q})$
of a spin-spiral described by means of the generalized
Bloch theorem and neglecting  spin-orbit coupling (SOC) \cite{MFE11} (full diamonds). 
In spite of the very different
approaches used, both curves are
close to each other over the  whole range of concentration considered.
Figure \ref{fig:FeNi_stiff} shows in addition experimental data obtained
using different techniques. As one can see, 
the calculations reproduce the experimental data fairly well.
For the  bcc as well as  fcc alloys, agreement between theory and
experiment is best in the regime of concentrated allloys 
and gets less satisfying  when 
approaching the Fe or Ni, respectively, rich regimes.

\begin{figure}[h]
\includegraphics[width=0.42\textwidth,angle=0,clip]{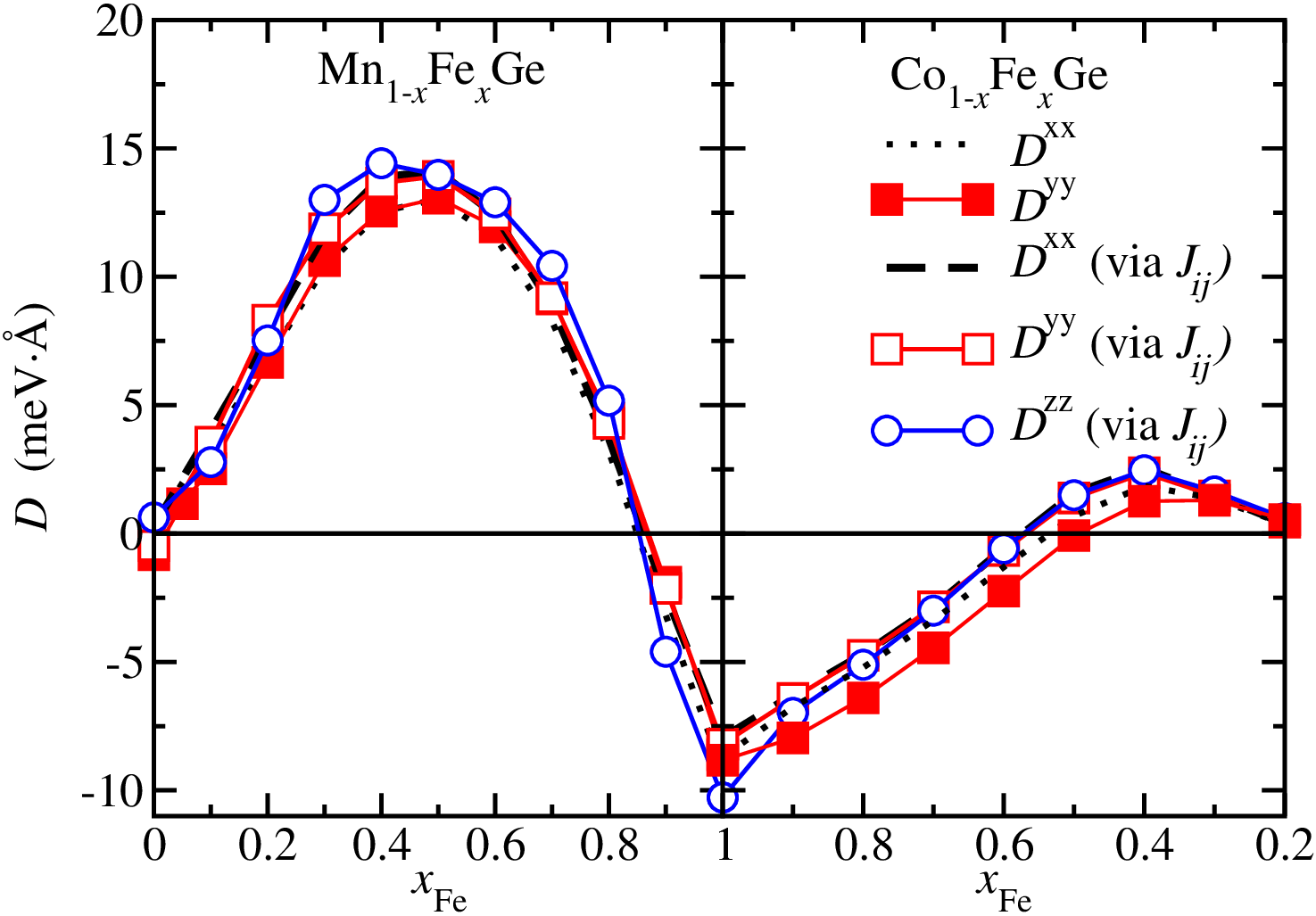}\;
\caption{\label{fig:MnGe_DMI} The 
  elements $D^{xx}$, $D^{yy}$ and $D^{zz}$
  of the micromagnetic DMI tensor calculated for
  Mn$_{1-x}$Fe$_{x}$Ge (left) and Co$_{1-x}$Fe$_{x}$Ge represented as
  a function of Fe concentration. }  
\end{figure}

In order to demonstrate the application 
of the derived expresssion for the  
micromagnetic DMI tensor, we consider here two different
non-centrosymmetric  system types, Mn$_{1-x}$Fe$_x$Ge and
Co$_{1-x}$Fe$_x$Ge, respectively,  having the  cubic B20 structure
 and the strongly anisotropic multilayer  system
(Cu/Fe$_{1-x}$Co$_x$/Pt)$_n$. 
Focusing first on the B20 systems, Fig. \ref{fig:MnGe_DMI} represents 
results for the three
diagonal elements $D^{xx}$, $D^{yy}$ and
$D^{zz}$ of the micromagnetic DMI tensor, 
with the elements   $D^{xx}$, $D^{yy}$  calculated
in two different ways.
%
%
In the first case, the magnetization has been assumed to be oriented 
along $\hat{z}$. The elements $D^{xx}$ and $D^{yy}$  
could  then be  calculated on the basis of  
Eq.\ (\ref{Eq:DMI_XY}) (dotted lines and full squares, respectively). 
In the second case, the expression
\begin{eqnarray}
 D^{\alpha\alpha}  &=&  \sum_{j \neq 0} D^\alpha_{0j} \, (\vec{R}_j - \vec{R}_0)_\alpha    
\label{Eq:DMI_micmag_vs_Heis}
\end{eqnarray}
based on the interatomic DMI elements ${D}^{x}_{ij}$  and ${D}^{y}_{ij}$
\cite{ME17} 
has been used  (dashed line and open squares, respectively)
 with the summation performed 
 for $|\vec{R}_i - \vec{R}_j| \le 6.5~a$, where $a$ is the lattice parameter.
 The difference between the two sets of  results can be
attributed first of all to the  restricted summation  in the
latter case.

The  element $D^{zz}$ has been calculated only using
Eq.\ (\ref{Eq:DMI_micmag_vs_Heis}) (open circles),  
with ${D}^{z}_{ij}$ obtained via an expression reported
previously \cite{EM09a} and having direct relation to Eq.\ (\ref{Eq_MM_DMI_zz}).
Although the consideres B20 systems have cubic Bravais lattice, the 
elements $D^{xx}$, $D^{yy}$ and $D^{zz}$ 
are not the same because of the reduced symmetry \cite{MWPE18}.
Nevertheless, the difference between all three components is rather small.
This is a rather important result for the systems under consideration since
the expression used for the calculation of the term $D^{zz}$ is rather
different from the one used for the two other terms.
It should be noted,  that the non-vanishing 
off-diagonal elements $D^{\mu\alpha}$ of the DMI tensor 
are substantially smaller  than the
diagonal elements \cite{MWPE18}  for the considered  systems.

\begin{figure}
\includegraphics[width=0.42\textwidth,angle=0,clip]{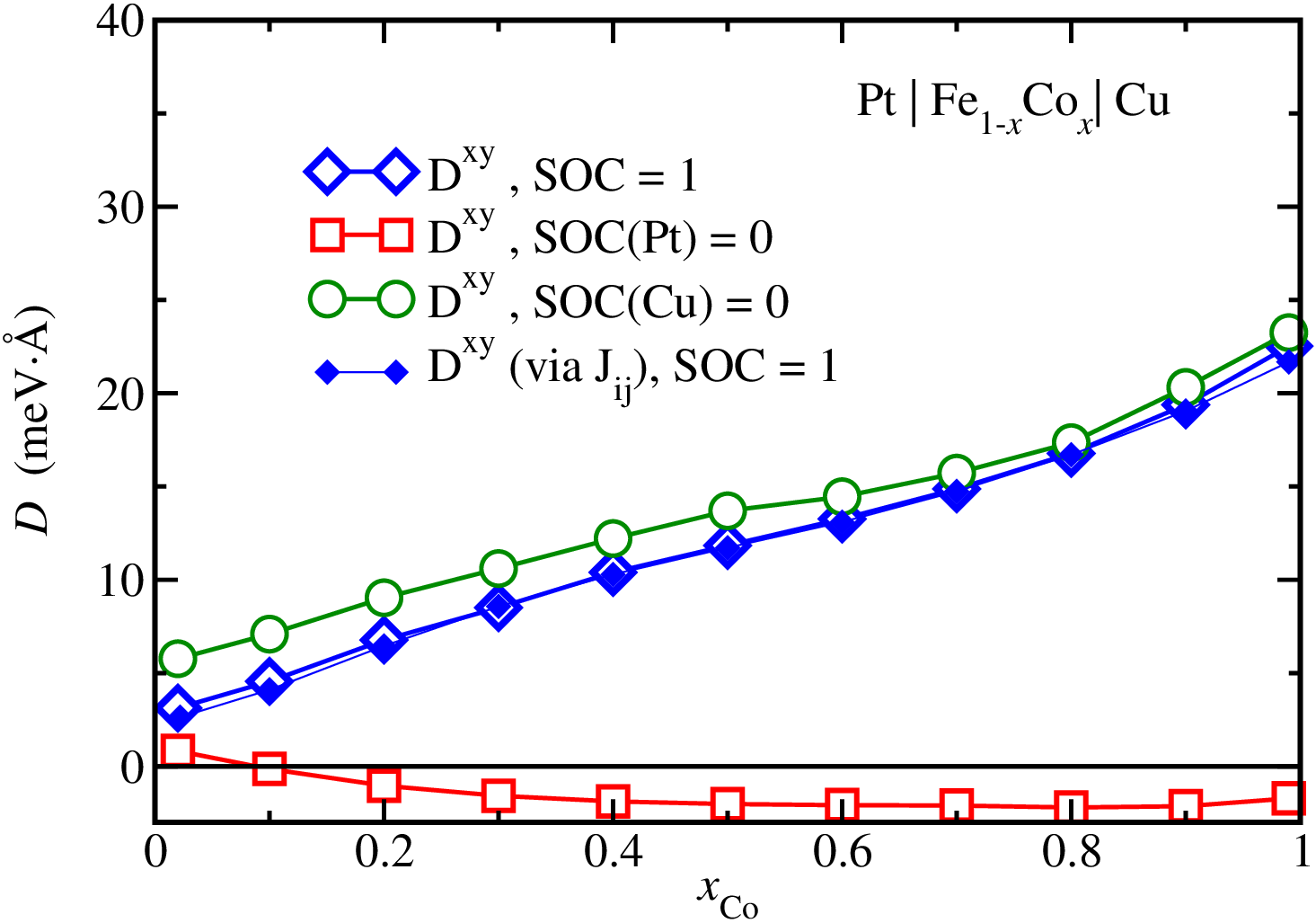}\;
\caption{\label{fig:FeCo_DMI}
  The   elements $D^{xy}$ of the micromagnetic DMI tensor
 for multilayer (Cu/Fe$_{1-x}$Co$_x$/Pt)$_n$ system as function of Co
 concentration.  }  
\end{figure}

In contrast to the discussed B20 alloys, 
the symmetry  of the (Cu/Fe$_{1-x}$Co$_x$/Pt)$_n$ multilayer
system results in a vanishing of the diagonal elements of the micromagnetic DMI
tensor and only the elements $D^{xy}$ and $D^{yx}$ 
 are non-zero for the magnetization
direction along $\hat{z}$, having opposite sign, $D^{yx} =
-D^{xy}$. This is in line with the symmetry properties 
of the Fermi-sea contribution to the spin-orbit torque (SOT) tensor
discussed previously by Wimmer et al.\  \cite{WCS+16}, that should be
obeyed also by the DMI
tensor  due to the relationship 
between these two tensors \cite{FBM14,MWPE18}.
The  element  $D^{xy}$ of the micromagnetic DMI tensor is plotted in
Fig.\ \ref{fig:FeCo_DMI} (open diamonds) as a function of Co
concentration, exhibiting a 
monotonous increase with the increase of Co concentration. 
%
Almost the
same behaviour is shown by the results obtained via Eq.\ (\ref{Eq:DMI_XY})
(full diamonds) with a small deviation caused by the cutoff in the summation
over the neighboring shells in this case.

Additional calculations have been performed to find out which 
atom type with its intrinsic SOC play the major role concerning
 the strength of the DMI in
the multi-component systems under consideration. 
In the case of B20 alloys the DMI strength is mainly determined by the
SOC of the $3d$-atoms. This follows immediately from a gradual decrease of
the DMI when the SOC of these atoms is scaled to be zero. 
The $p$-states of Ge in the B20 materials are strongly hybridized with
the $d$-states of the $3d$-atoms and therefore have a
key role in mediating the antisymmetric exchange interactions, as was
previously discussed in the literature (see, e.g.\ \cite{GFS+15}).
In particular, the dependence of the DMI on the 
relative position of the $p$-states of Ge and
 $d$-states of the transition metals with
respect to each other as well as with respect to the Fermi energy,
 leading to a sign change of the DMI parameters upon
 variation of the composition in these alloys,
 has been demonstrated \cite{GFS+15}.

In the case of the  multilayer system (Cu/Fe$_{1-x}$Co$_x$/Pt)$_n$,
switching off the SOC for the $3d$-atoms does not result in a
significant change of the DMI. A similar result is found when the SOC for the Cu
atoms is switched off (open circles); i.e.\ a weak increase of the
DMI is seen almost over all the concentration region. However, the
magnitude of the components $D^{xy}$ and $D^{yx}$ drops down when 
the SOC of the Pt atoms (open squares) is switched off
 and  $D^{xy}$ even changes sign at $x \approx 0.1$. 
Obviously, one reason leading to this behaviour is the  
strong    SOC for the  Pt atoms.

\begin{figure}
\includegraphics[width=0.4\textwidth,angle=0,clip]{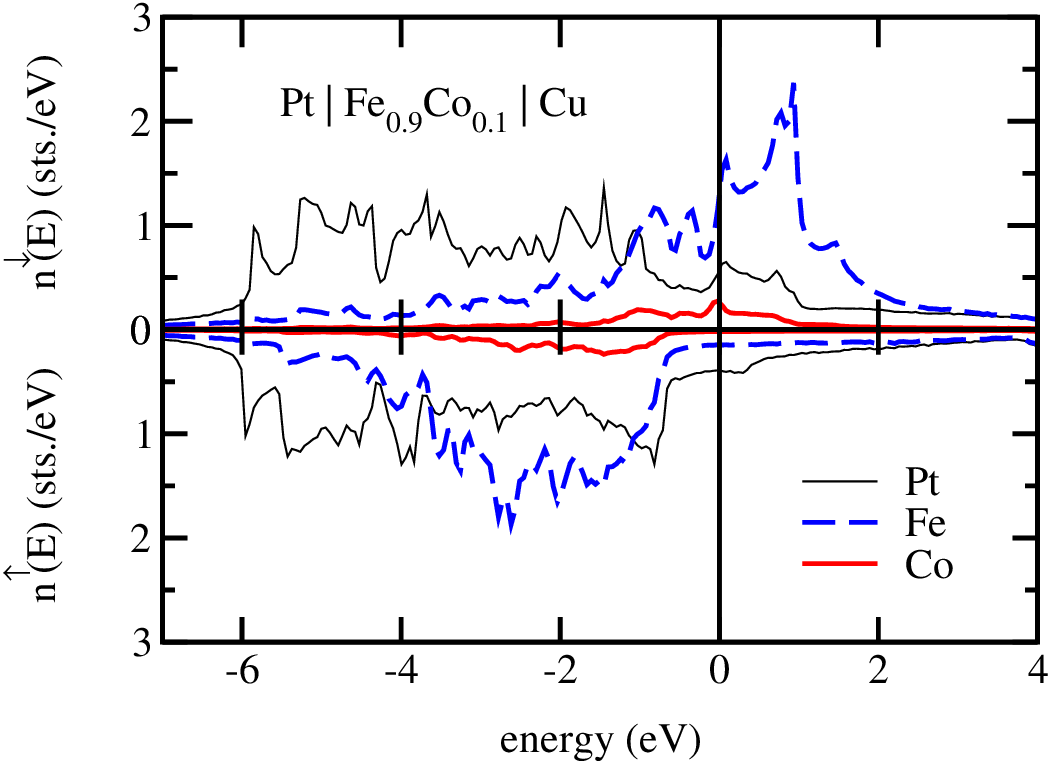}\;(a)
\includegraphics[width=0.4\textwidth,angle=0,clip]{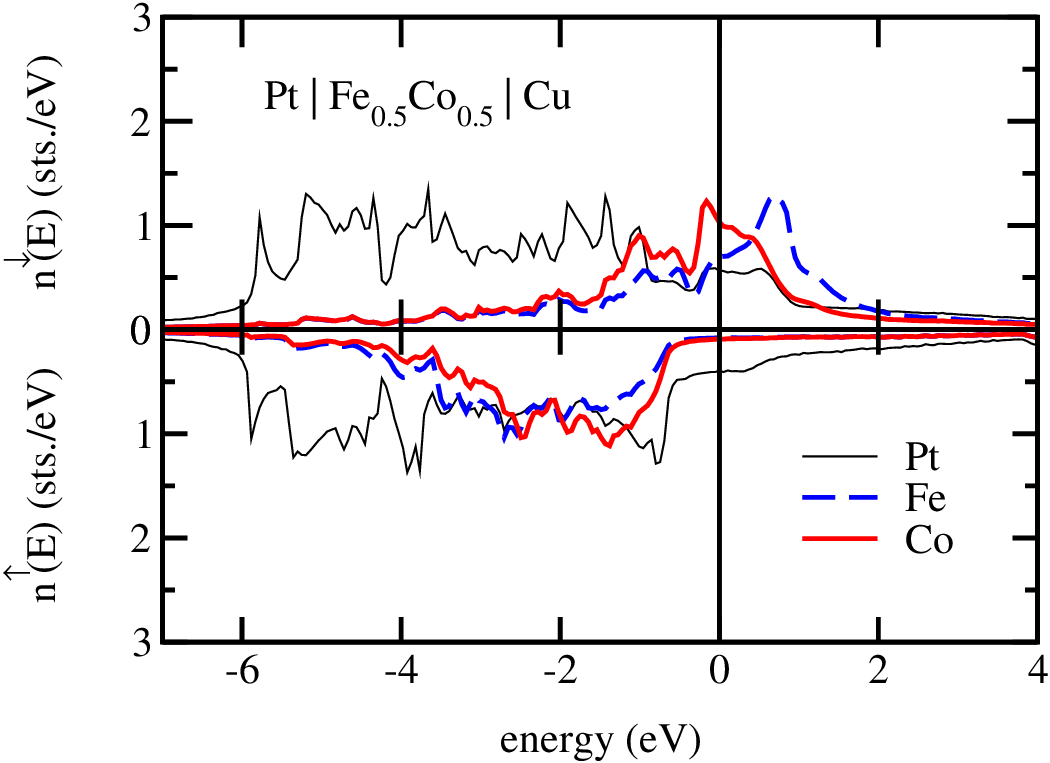}\;(b)
\includegraphics[width=0.4\textwidth,angle=0,clip]{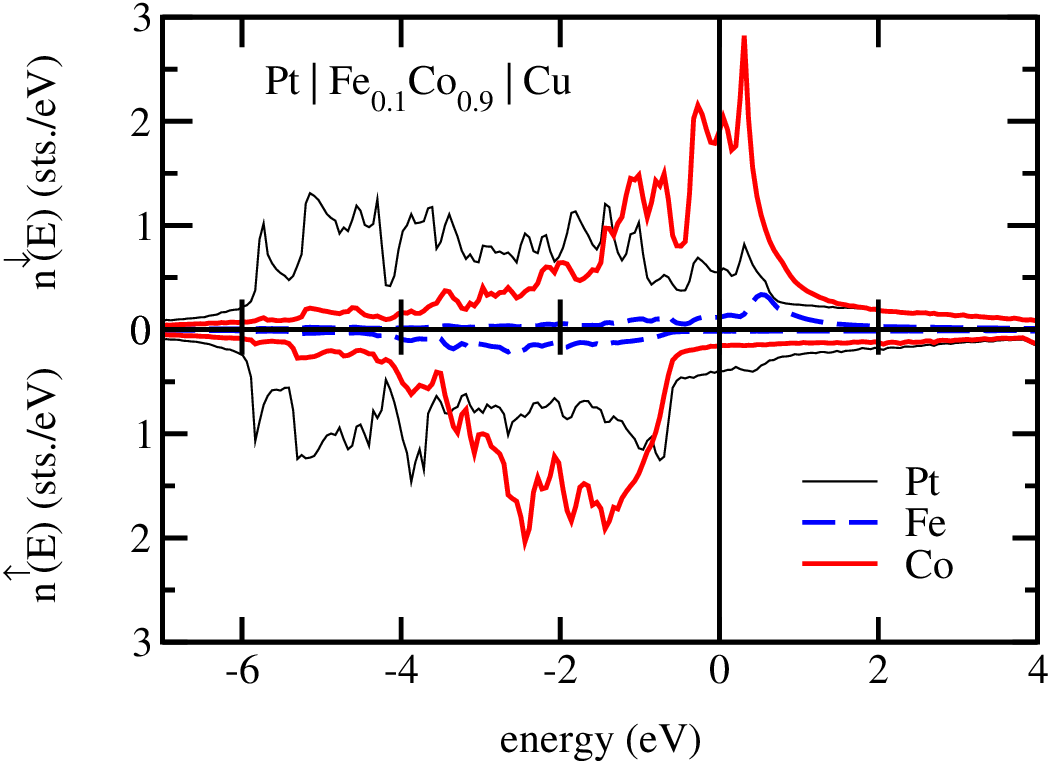}\;(c)
\includegraphics[width=0.4\textwidth,angle=0,clip]{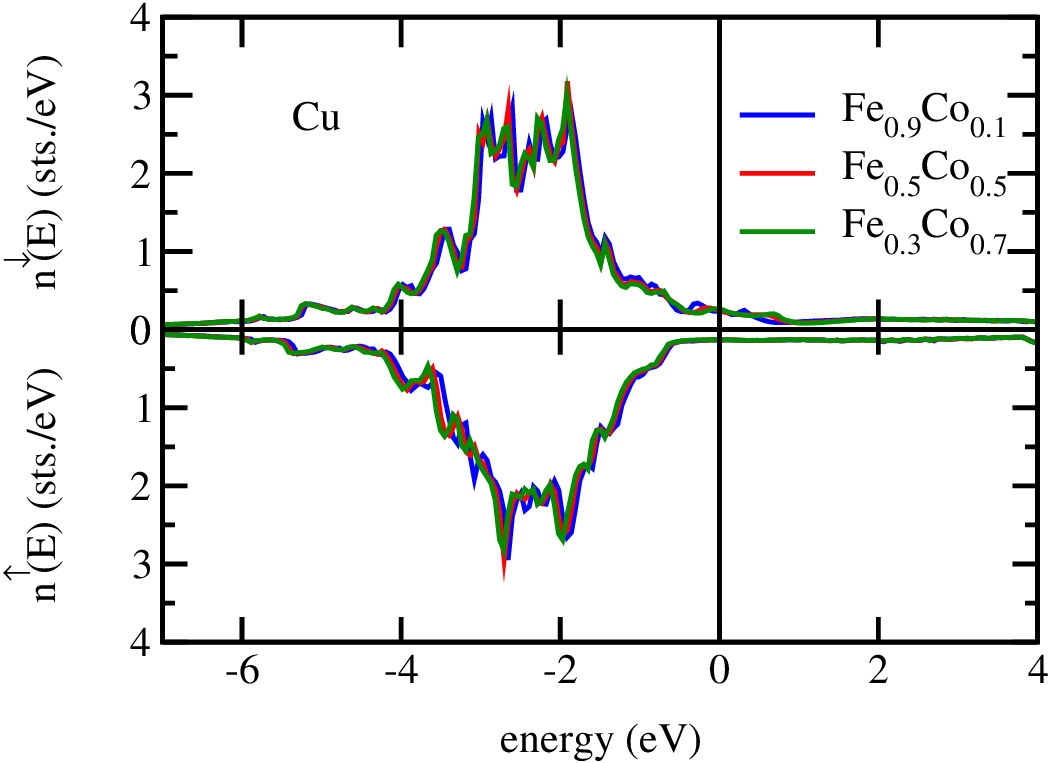}\;(d)
 \caption{\label{fig:FeCo_DOS} Spin- and element resolved $d$-DOS for the
   $d$-states in the multilayer  system (Cu/Fe$_{1-x}$Co$_x$/Pt)$_n$
     for $x = 0.1$ (a), $x = 0.5$ (b) and  $x = 0.9$ (c):  Pt
   (black line), Co (red line) and Fe (blue line).
  The  $d$-DOS for the 
   Cu atoms in the systems  are shown
 in (d) for three different concentrations.}   
 \end{figure}

To allow for a more detailed discussion, 
we present in Fig.\ \ref{fig:FeCo_DOS} the element
resolved density of states (DOS) of (Cu/Fe$_{1-x}$Co$_x$/Pt)$_n$.
%
 Fig.\ \ref{fig:FeCo_DOS}(d) shows that the Cu $d$-bands
lie well below the Fermi energy. The fact that they are rather narrow
indicates their relatively weak 
hybridization with the Fe and Co states. As a consequence this results
in a weak modification of the $d$-states of Cu, when the magnetic
moments of the $3d$-atoms within the (Fe,Co) layer are rotated to form 
a spin spiral in the system. In addition, one can say that
the SOC, which is responsible for the antisymmetric exchange,
 is rather small for the  $3d$-states of the Cu atoms. 
Thus, both properties lead to a small contribution 
of the Cu layer to the DMI strength. 
On the other side, the heavy Pt atoms are characterized by a strong SOC.
The partially occupied Pt $d$-bands are rather broad and substantially
overlap in energy with the Co and Fe energy bands. This results (see Fig.\
\ref{fig:FeCo_DOS}) in a strong spin-dependent hybridization of the Pt states 
with the states of Co (more pronounced) and 
Fe (less pronounced) that in turn leads to a stronger modification of the Pt
states due to a tilting of the spin moments in the (Co,Fe) layer; in particular
upon creation of a spin spiral.
The lack of inversion symmetry results in different
changes of the SOC-induced anticrossing gaps associated with spin
spirals having different spin helicity, that  at the end determines the sign
and magnitude of the DMI vector (see \cite{San17}). 
A larger exchange splitting and, as a result, larger spin moment of the Co
atoms can be responsible for a stronger DMI in the Co-rich limit due to a
stronger perturbation caused in the neighboring layers upon rotation
of the Co magnetic moments. 

\section{Summary}

In summary, an expression for the spin-stiffness parameter
has been worked out within the
fully relativistic KKR Green function method  that can be seen as a
relativistic extension of the non-relativistic expression reported
previously in the literature. The approach 
gives  also access to calculations 
of the micromagnetic DMI tensor with all elements calculated for 
a feromagnetic orientation of the magnetization and a fixed frame of
references. The expression for the  element $D^{z\alpha}$
of the DMI tensor differs from the expressions for 
the elements $D^{x\alpha}$ and  $D^{y\alpha}$  as the former 
is derived via second-order perturbation theory concerning the 
the energy change caused by formation of a  spin-spiral 
while the latter elements can be associated with
the first-order term.  The first applications of the 
presented expression lead to a rather good agreement 
with available experimental data.

 \section{Acknowledgement}
Financial support by the DFG via SFB 1277 (Emergent Relativistic Effects
in Condensed Matter - From Fundamental Aspects to Electronic
Functionality), as well as the DFG financial support
via the priority programs EB154/36-1 are gratefully acknowledged.


\begin{thebibliography}{26}%
\makeatletter
\providecommand \@ifxundefined [1]{%
 \@ifx{#1\undefined}
}%
\providecommand \@ifnum [1]{%
 \ifnum #1\expandafter \@firstoftwo
 \else \expandafter \@secondoftwo
 \fi
}%
\providecommand \@ifx [1]{%
 \ifx #1\expandafter \@firstoftwo
 \else \expandafter \@secondoftwo
 \fi
}%
\providecommand \natexlab [1]{#1}%
\providecommand \enquote  [1]{``#1''}%
\providecommand \bibnamefont  [1]{#1}%
\providecommand \bibfnamefont [1]{#1}%
\providecommand \citenamefont [1]{#1}%
\providecommand \href@noop [0]{\@secondoftwo}%
\providecommand \href [0]{\begingroup \@sanitize@url \@href}%
\providecommand \@href[1]{\@@startlink{#1}\@@href}%
\providecommand \@@href[1]{\endgroup#1\@@endlink}%
\providecommand \@sanitize@url [0]{\catcode `\\12\catcode `\$12\catcode
  `\&12\catcode `\#12\catcode `\^12\catcode `\_12\catcode `\%12\relax}%
\providecommand \@@startlink[1]{}%
\providecommand \@@endlink[0]{}%
\providecommand \url  [0]{\begingroup\@sanitize@url \@url }%
\providecommand \@url [1]{\endgroup\@href {#1}{\urlprefix }}%
\providecommand \urlprefix  [0]{URL }%
\providecommand \Eprint [0]{\href }%
\providecommand \doibase [0]{http://dx.doi.org/}%
\providecommand \selectlanguage [0]{\@gobble}%
\providecommand \bibinfo  [0]{\@secondoftwo}%
\providecommand \bibfield  [0]{\@secondoftwo}%
\providecommand \translation [1]{[#1]}%
\providecommand \BibitemOpen [0]{}%
\providecommand \bibitemStop [0]{}%
\providecommand \bibitemNoStop [0]{.\EOS\space}%
\providecommand \EOS [0]{\spacefactor3000\relax}%
\providecommand \BibitemShut  [1]{\csname bibitem#1\endcsname}%
\let\auto@bib@innerbib\@empty
\bibitem [{\citenamefont {Liechtenstein}\ \emph {et~al.}(1987)\citenamefont
  {Liechtenstein}, \citenamefont {Katsnelson}, \citenamefont {Antropov},\ and\
  \citenamefont {Gubanov}}]{LKAG87}%
  \BibitemOpen
  \bibfield  {author} {\bibinfo {author} {\bibfnamefont {A.~I.}\ \bibnamefont
  {Liechtenstein}}, \bibinfo {author} {\bibfnamefont {M.~I.}\ \bibnamefont
  {Katsnelson}}, \bibinfo {author} {\bibfnamefont {V.~P.}\ \bibnamefont
  {Antropov}}, \ and\ \bibinfo {author} {\bibfnamefont {V.~A.}\ \bibnamefont
  {Gubanov}},\ }\href {\doibase 10.1016/0304-8853(87)90721-9} {\bibfield
  {journal} {\bibinfo  {journal} {J. Magn. Magn. Materials}\ }\textbf {\bibinfo
  {volume} {67}},\ \bibinfo {pages} {65} (\bibinfo {year} {1987})}\BibitemShut
  {NoStop}%
\bibitem [{\citenamefont {Antropov}\ \emph {et~al.}(1997)\citenamefont
  {Antropov}, \citenamefont {Katsnelson},\ and\ \citenamefont
  {Liechtenstein}}]{AKL97}%
  \BibitemOpen
  \bibfield  {author} {\bibinfo {author} {\bibfnamefont {V.}~\bibnamefont
  {Antropov}}, \bibinfo {author} {\bibfnamefont {M.}~\bibnamefont
  {Katsnelson}}, \ and\ \bibinfo {author} {\bibfnamefont {A.}~\bibnamefont
  {Liechtenstein}},\ }\href {\doibase
  https://doi.org/10.1016/S0921-4526(97)00203-2} {\bibfield  {journal}
  {\bibinfo  {journal} {Physica B: Condensed Matter}\ }\textbf {\bibinfo
  {volume} {237-238}},\ \bibinfo {pages} {336 } (\bibinfo {year} {1997})},\
  \bibinfo {note} {proceedings of the Yamada Conference XLV, the International
  Conference on the Physics of Transition Metals}\BibitemShut {NoStop}%
\bibitem [{\citenamefont {K\"ubler}(2009)}]{Kub09}%
  \BibitemOpen
  \bibfield  {author} {\bibinfo {author} {\bibfnamefont {J.}~\bibnamefont
  {K\"ubler}},\ }\href {https://books.google.de/books?id=ZbM0gHCcmaQC} {\emph
  {\bibinfo {title} {Theory of Itinerant Electron Magnetism}}},\ International
  Series of Monographs on Physics\ (\bibinfo  {publisher} {OUP Oxford},\
  \bibinfo {year} {2009})\BibitemShut {NoStop}%
\bibitem [{\citenamefont {Udvardi}\ \emph {et~al.}(2003)\citenamefont
  {Udvardi}, \citenamefont {Szunyogh}, \citenamefont {Palot\'as},\ and\
  \citenamefont {Weinberger}}]{USPW03}%
  \BibitemOpen
  \bibfield  {author} {\bibinfo {author} {\bibfnamefont {L.}~\bibnamefont
  {Udvardi}}, \bibinfo {author} {\bibfnamefont {L.}~\bibnamefont {Szunyogh}},
  \bibinfo {author} {\bibfnamefont {K.}~\bibnamefont {Palot\'as}}, \ and\
  \bibinfo {author} {\bibfnamefont {P.}~\bibnamefont {Weinberger}},\ }\href
  {\doibase 10.1103/PhysRevB.68.104436} {\bibfield  {journal} {\bibinfo
  {journal} {Phys. Rev. B}\ }\textbf {\bibinfo {volume} {68}},\ \bibinfo
  {pages} {104436} (\bibinfo {year} {2003})}\BibitemShut {NoStop}%
\bibitem [{\citenamefont {Ebert}\ and\ \citenamefont
  {Mankovsky}(2009)}]{EM09a}%
  \BibitemOpen
  \bibfield  {author} {\bibinfo {author} {\bibfnamefont {H.}~\bibnamefont
  {Ebert}}\ and\ \bibinfo {author} {\bibfnamefont {S.}~\bibnamefont
  {Mankovsky}},\ }\href {\doibase 10.1103/PhysRevB.79.045209} {\bibfield
  {journal} {\bibinfo  {journal} {Phys. Rev. B}\ }\textbf {\bibinfo {volume}
  {79}},\ \bibinfo {pages} {045209} (\bibinfo {year} {2009})}\BibitemShut
  {NoStop}%
\bibitem [{\citenamefont {Zeller}(2008)}]{Zel08}%
  \BibitemOpen
  \bibfield  {author} {\bibinfo {author} {\bibfnamefont {R.}~\bibnamefont
  {Zeller}},\ }\href {\doibase 10.1088/0953-8984/20/29/294215} {\bibfield
  {journal} {\bibinfo  {journal} {J. Phys.: Cond. Mat.}\ }\textbf {\bibinfo
  {volume} {20}},\ \bibinfo {pages} {294215} (\bibinfo {year}
  {2008})}\BibitemShut {NoStop}%
\bibitem [{\citenamefont {Bogdanov}\ and\ \citenamefont {Hubert}(1994)}]{BH94}%
  \BibitemOpen
  \bibfield  {author} {\bibinfo {author} {\bibfnamefont {A.}~\bibnamefont
  {Bogdanov}}\ and\ \bibinfo {author} {\bibfnamefont {A.}~\bibnamefont
  {Hubert}},\ }\href {\doibase http://dx.doi.org/10.1016/0304-8853(94)90046-9}
  {\bibfield  {journal} {\bibinfo  {journal} {J. Magn. Magn. Materials}\
  }\textbf {\bibinfo {volume} {138}},\ \bibinfo {pages} {255 } (\bibinfo {year}
  {1994})}\BibitemShut {NoStop}%
\bibitem [{\citenamefont {K{\"u}bler}(2000)}]{Kueb00}%
  \BibitemOpen
  \bibfield  {author} {\bibinfo {author} {\bibfnamefont {J.}~\bibnamefont
  {K{\"u}bler}},\ }\enquote {\bibinfo {title} {Theory of itinerant electron
  magnetism},}\ \ (\bibinfo  {publisher} {Oxford University Press},\ \bibinfo
  {address} {Oxford},\ \bibinfo {year} {2000})\ p.\ \bibinfo {pages}
  {460}\BibitemShut {NoStop}%
\bibitem [{\citenamefont {Hamrle}\ \emph {et~al.}(2009)\citenamefont {Hamrle},
  \citenamefont {Gaier}, \citenamefont {Min}, \citenamefont {Hillebrands},
  \citenamefont {Sakuraba},\ and\ \citenamefont {Ando}}]{HGM+09}%
  \BibitemOpen
  \bibfield  {author} {\bibinfo {author} {\bibfnamefont {J.}~\bibnamefont
  {Hamrle}}, \bibinfo {author} {\bibfnamefont {O.}~\bibnamefont {Gaier}},
  \bibinfo {author} {\bibfnamefont {S.-G.}\ \bibnamefont {Min}}, \bibinfo
  {author} {\bibfnamefont {B.}~\bibnamefont {Hillebrands}}, \bibinfo {author}
  {\bibfnamefont {Y.}~\bibnamefont {Sakuraba}}, \ and\ \bibinfo {author}
  {\bibfnamefont {Y.}~\bibnamefont {Ando}},\ }\href
  {http://stacks.iop.org/0022-3727/42/i=8/a=084005} {\bibfield  {journal}
  {\bibinfo  {journal} {Journal of Physics D: Applied Physics}\ }\textbf
  {\bibinfo {volume} {42}},\ \bibinfo {pages} {084005} (\bibinfo {year}
  {2009})}\BibitemShut {NoStop}%
\bibitem [{\citenamefont {Liechtenstein}\ \emph {et~al.}(1984)\citenamefont
  {Liechtenstein}, \citenamefont {Katsnelson},\ and\ \citenamefont
  {Gubanov}}]{LKG84}%
  \BibitemOpen
  \bibfield  {author} {\bibinfo {author} {\bibfnamefont {A.~I.}\ \bibnamefont
  {Liechtenstein}}, \bibinfo {author} {\bibfnamefont {M.~I.}\ \bibnamefont
  {Katsnelson}}, \ and\ \bibinfo {author} {\bibfnamefont {V.~A.}\ \bibnamefont
  {Gubanov}},\ }\href {\doibase 10.1088/0305-4608/14/7/007} {\bibfield
  {journal} {\bibinfo  {journal} {J. Phys. F: Met. Phys.}\ }\textbf {\bibinfo
  {volume} {14}},\ \bibinfo {pages} {L125} (\bibinfo {year}
  {1984})}\BibitemShut {NoStop}%
\bibitem [{\citenamefont {{Freimuth}}\ \emph {et~al.}(2014)\citenamefont
  {{Freimuth}}, \citenamefont {{Bl{\"u}gel}},\ and\ \citenamefont
  {{Mokrousov}}}]{FBM14}%
  \BibitemOpen
  \bibfield  {author} {\bibinfo {author} {\bibfnamefont {F.}~\bibnamefont
  {{Freimuth}}}, \bibinfo {author} {\bibfnamefont {S.}~\bibnamefont
  {{Bl{\"u}gel}}}, \ and\ \bibinfo {author} {\bibfnamefont {Y.}~\bibnamefont
  {{Mokrousov}}},\ }\href {\doibase 10.1088/0953-8984/26/10/104202} {\bibfield
  {journal} {\bibinfo  {journal} {J. Phys.: Cond. Mat.}\ }\textbf {\bibinfo
  {volume} {26}},\ \bibinfo {pages} {104202} (\bibinfo {year} {2014})},\
  \Eprint {http://arxiv.org/abs/1308.5983} {arXiv:1308.5983} \BibitemShut
  {NoStop}%
\bibitem [{\citenamefont {MacDonald}\ and\ \citenamefont {Vosko}(1979)}]{MV79}%
  \BibitemOpen
  \bibfield  {author} {\bibinfo {author} {\bibfnamefont {A.~H.}\ \bibnamefont
  {MacDonald}}\ and\ \bibinfo {author} {\bibfnamefont {S.~H.}\ \bibnamefont
  {Vosko}},\ }\href {\doibase 10.1088/0022-3719/12/15/007} {\bibfield
  {journal} {\bibinfo  {journal} {J. Phys. C: Solid State Phys.}\ }\textbf
  {\bibinfo {volume} {12}},\ \bibinfo {pages} {2977} (\bibinfo {year}
  {1979})}\BibitemShut {NoStop}%
\bibitem [{\citenamefont {Rose}(1961)}]{Ros61}%
  \BibitemOpen
  \bibfield  {author} {\bibinfo {author} {\bibfnamefont {M.~E.}\ \bibnamefont
  {Rose}},\ }\href
  {http://openlibrary.org/works/OL3517103W/Relativistic_electron_theory} {\emph
  {\bibinfo {title} {Relativistic Electron Theory}}}\ (\bibinfo  {publisher}
  {Wiley},\ \bibinfo {address} {New York},\ \bibinfo {year} {1961})\BibitemShut
  {NoStop}%
\bibitem [{\citenamefont {Ebert}\ \emph {et~al.}(2016)\citenamefont {Ebert},
  \citenamefont {Braun}, \citenamefont {K\"odderitzsch},\ and\ \citenamefont
  {Mankovsky}}]{EBKM16}%
  \BibitemOpen
  \bibfield  {author} {\bibinfo {author} {\bibfnamefont {H.}~\bibnamefont
  {Ebert}}, \bibinfo {author} {\bibfnamefont {J.}~\bibnamefont {Braun}},
  \bibinfo {author} {\bibfnamefont {D.}~\bibnamefont {K\"odderitzsch}}, \ and\
  \bibinfo {author} {\bibfnamefont {S.}~\bibnamefont {Mankovsky}},\ }\href
  {\doibase 10.1103/PhysRevB.93.075145} {\bibfield  {journal} {\bibinfo
  {journal} {Phys. Rev. B}\ }\textbf {\bibinfo {volume} {93}},\ \bibinfo
  {pages} {075145} (\bibinfo {year} {2016})}\BibitemShut {NoStop}%
\bibitem [{\citenamefont {Mankovsky}\ and\ \citenamefont
  {Ebert}(2017{\natexlab{a}})}]{MWE18}%
  \BibitemOpen
  \bibfield  {author} {\bibinfo {author} {\bibfnamefont {S.}~\bibnamefont
  {Mankovsky}}\ and\ \bibinfo {author} {\bibfnamefont {H.}~\bibnamefont
  {Ebert}},\ }\href {\doibase 10.1103/PhysRevB.96.104416} {\bibfield  {journal}
  {\bibinfo  {journal} {Phys. Rev. B}\ }\textbf {\bibinfo {volume} {96}},\
  \bibinfo {pages} {104416} (\bibinfo {year} {2017}{\natexlab{a}})}\BibitemShut
  {NoStop}%
\bibitem [{\citenamefont {Mankovsky}\ and\ \citenamefont
  {Ebert}(2017{\natexlab{b}})}]{ME17}%
  \BibitemOpen
  \bibfield  {author} {\bibinfo {author} {\bibfnamefont {S.}~\bibnamefont
  {Mankovsky}}\ and\ \bibinfo {author} {\bibfnamefont {H.}~\bibnamefont
  {Ebert}},\ }\href {\doibase 10.1103/PhysRevB.96.104416} {\bibfield  {journal}
  {\bibinfo  {journal} {Phys. Rev. B}\ }\textbf {\bibinfo {volume} {96}},\
  \bibinfo {pages} {104416} (\bibinfo {year} {2017}{\natexlab{b}})}\BibitemShut
  {NoStop}%
\bibitem [{\citenamefont {Koretsune}\ \emph {et~al.}(2015)\citenamefont
  {Koretsune}, \citenamefont {Nagaosa},\ and\ \citenamefont {Arita}}]{KNA15}%
  \BibitemOpen
  \bibfield  {author} {\bibinfo {author} {\bibfnamefont {T.}~\bibnamefont
  {Koretsune}}, \bibinfo {author} {\bibfnamefont {N.}~\bibnamefont {Nagaosa}},
  \ and\ \bibinfo {author} {\bibfnamefont {R.}~\bibnamefont {Arita}},\ }\href
  {http://dx.doi.org/10.1038/srep13302} {\bibfield  {journal} {\bibinfo
  {journal} {Scientific Reports}\ }\textbf {\bibinfo {volume} {5}},\ \bibinfo
  {pages} {13302} (\bibinfo {year} {2015})}\BibitemShut {NoStop}%
\bibitem [{\citenamefont {Mankovsky}\ \emph {et~al.}(2009)\citenamefont
  {Mankovsky}, \citenamefont {Bornemann}, \citenamefont {Min\'ar},
  \citenamefont {Polesya}, \citenamefont {Ebert}, \citenamefont {Staunton},\
  and\ \citenamefont {Lichtenstein}}]{MBM+09}%
  \BibitemOpen
  \bibfield  {author} {\bibinfo {author} {\bibfnamefont {S.}~\bibnamefont
  {Mankovsky}}, \bibinfo {author} {\bibfnamefont {S.}~\bibnamefont
  {Bornemann}}, \bibinfo {author} {\bibfnamefont {J.}~\bibnamefont {Min\'ar}},
  \bibinfo {author} {\bibfnamefont {S.}~\bibnamefont {Polesya}}, \bibinfo
  {author} {\bibfnamefont {H.}~\bibnamefont {Ebert}}, \bibinfo {author}
  {\bibfnamefont {J.~B.}\ \bibnamefont {Staunton}}, \ and\ \bibinfo {author}
  {\bibfnamefont {A.~I.}\ \bibnamefont {Lichtenstein}},\ }\href {\doibase
  10.1103/PhysRevB.80.014422} {\bibfield  {journal} {\bibinfo  {journal} {Phys.
  Rev. B}\ }\textbf {\bibinfo {volume} {80}},\ \bibinfo {pages} {014422}
  (\bibinfo {year} {2009})}\BibitemShut {NoStop}%
\bibitem [{\citenamefont {Nakai}(1983)}]{Nak83}%
  \BibitemOpen
  \bibfield  {author} {\bibinfo {author} {\bibfnamefont {I.}~\bibnamefont
  {Nakai}},\ }\href@noop {} {\bibfield  {journal} {\bibinfo  {journal} {J.
  Phys. Soc. Japan}\ }\textbf {\bibinfo {volume} {52}},\ \bibinfo {pages}
  {1781} (\bibinfo {year} {1983})}\BibitemShut {NoStop}%
\bibitem [{\citenamefont {Hatherly}\ \emph {et~al.}(1964)\citenamefont
  {Hatherly}, \citenamefont {Hirakawa}, \citenamefont {Lowde}, \citenamefont
  {Mallett}, \citenamefont {Stringfellow},\ and\ \citenamefont
  {Torrie}}]{HHL+64}%
  \BibitemOpen
  \bibfield  {author} {\bibinfo {author} {\bibfnamefont {M.}~\bibnamefont
  {Hatherly}}, \bibinfo {author} {\bibfnamefont {K.}~\bibnamefont {Hirakawa}},
  \bibinfo {author} {\bibfnamefont {R.~D.}\ \bibnamefont {Lowde}}, \bibinfo
  {author} {\bibfnamefont {J.~F.}\ \bibnamefont {Mallett}}, \bibinfo {author}
  {\bibfnamefont {M.~W.}\ \bibnamefont {Stringfellow}}, \ and\ \bibinfo
  {author} {\bibfnamefont {B.~H.}\ \bibnamefont {Torrie}},\ }\href@noop {}
  {\bibfield  {journal} {\bibinfo  {journal} {Proc. Phys. Soc. (London)}\
  }\textbf {\bibinfo {volume} {84}},\ \bibinfo {pages} {55} (\bibinfo {year}
  {1964})}\BibitemShut {NoStop}%
\bibitem [{\citenamefont {Rusov}(1967)}]{Rus76}%
  \BibitemOpen
  \bibfield  {author} {\bibinfo {author} {\bibfnamefont {G.~I.}\ \bibnamefont
  {Rusov}},\ }\href@noop {} {\bibfield  {journal} {\bibinfo  {journal} {Sov.
  Phys.-Solid State}\ }\textbf {\bibinfo {volume} {9}},\ \bibinfo {pages} {146}
  (\bibinfo {year} {1967})}\BibitemShut {NoStop}%
\bibitem [{\citenamefont {Mankovsky}\ \emph {et~al.}(2011)\citenamefont
  {Mankovsky}, \citenamefont {Fecher},\ and\ \citenamefont {Ebert}}]{MFE11}%
  \BibitemOpen
  \bibfield  {author} {\bibinfo {author} {\bibfnamefont {S.}~\bibnamefont
  {Mankovsky}}, \bibinfo {author} {\bibfnamefont {G.~H.}\ \bibnamefont
  {Fecher}}, \ and\ \bibinfo {author} {\bibfnamefont {H.}~\bibnamefont
  {Ebert}},\ }\href {\doibase 10.1103/PhysRevB.83.144401} {\bibfield  {journal}
  {\bibinfo  {journal} {Phys. Rev. B}\ }\textbf {\bibinfo {volume} {83}},\
  \bibinfo {pages} {144401} (\bibinfo {year} {2011})}\BibitemShut {NoStop}%
\bibitem [{\citenamefont {Mankovsky}\ \emph {et~al.}(2018)\citenamefont
  {Mankovsky}, \citenamefont {Wimmer}, \citenamefont {Polesya},\ and\
  \citenamefont {Ebert}}]{MWPE18}%
  \BibitemOpen
  \bibfield  {author} {\bibinfo {author} {\bibfnamefont {S.}~\bibnamefont
  {Mankovsky}}, \bibinfo {author} {\bibfnamefont {S.}~\bibnamefont {Wimmer}},
  \bibinfo {author} {\bibfnamefont {S.}~\bibnamefont {Polesya}}, \ and\
  \bibinfo {author} {\bibfnamefont {H.}~\bibnamefont {Ebert}},\ }\href
  {\doibase 10.1103/PhysRevB.97.024403} {\bibfield  {journal} {\bibinfo
  {journal} {Phys. Rev. B}\ }\textbf {\bibinfo {volume} {97}},\ \bibinfo
  {pages} {024403} (\bibinfo {year} {2018})}\BibitemShut {NoStop}%
\bibitem [{\citenamefont {Wimmer}\ \emph {et~al.}(2016)\citenamefont {Wimmer},
  \citenamefont {Chadova}, \citenamefont {Seemann}, \citenamefont
  {K\"odderitzsch},\ and\ \citenamefont {Ebert}}]{WCS+16}%
  \BibitemOpen
  \bibfield  {author} {\bibinfo {author} {\bibfnamefont {S.}~\bibnamefont
  {Wimmer}}, \bibinfo {author} {\bibfnamefont {K.}~\bibnamefont {Chadova}},
  \bibinfo {author} {\bibfnamefont {M.}~\bibnamefont {Seemann}}, \bibinfo
  {author} {\bibfnamefont {D.}~\bibnamefont {K\"odderitzsch}}, \ and\ \bibinfo
  {author} {\bibfnamefont {H.}~\bibnamefont {Ebert}},\ }\href {\doibase
  10.1103/PhysRevB.94.054415} {\bibfield  {journal} {\bibinfo  {journal} {Phys.
  Rev. B}\ }\textbf {\bibinfo {volume} {94}},\ \bibinfo {pages} {054415}
  (\bibinfo {year} {2016})}\BibitemShut {NoStop}%
\bibitem [{\citenamefont {Gayles}\ \emph {et~al.}(2015)\citenamefont {Gayles},
  \citenamefont {Freimuth}, \citenamefont {Schena}, \citenamefont {Lani},
  \citenamefont {Mavropoulos}, \citenamefont {Duine}, \citenamefont {Bl\"ugel},
  \citenamefont {Sinova},\ and\ \citenamefont {Mokrousov}}]{GFS+15}%
  \BibitemOpen
  \bibfield  {author} {\bibinfo {author} {\bibfnamefont {J.}~\bibnamefont
  {Gayles}}, \bibinfo {author} {\bibfnamefont {F.}~\bibnamefont {Freimuth}},
  \bibinfo {author} {\bibfnamefont {T.}~\bibnamefont {Schena}}, \bibinfo
  {author} {\bibfnamefont {G.}~\bibnamefont {Lani}}, \bibinfo {author}
  {\bibfnamefont {P.}~\bibnamefont {Mavropoulos}}, \bibinfo {author}
  {\bibfnamefont {R.~A.}\ \bibnamefont {Duine}}, \bibinfo {author}
  {\bibfnamefont {S.}~\bibnamefont {Bl\"ugel}}, \bibinfo {author}
  {\bibfnamefont {J.}~\bibnamefont {Sinova}}, \ and\ \bibinfo {author}
  {\bibfnamefont {Y.}~\bibnamefont {Mokrousov}},\ }\href {\doibase
  10.1103/PhysRevLett.115.036602} {\bibfield  {journal} {\bibinfo  {journal}
  {Phys. Rev. Lett.}\ }\textbf {\bibinfo {volume} {115}},\ \bibinfo {pages}
  {036602} (\bibinfo {year} {2015})}\BibitemShut {NoStop}%
\bibitem [{\citenamefont {Sandratskii}(2017)}]{San17}%
  \BibitemOpen
  \bibfield  {author} {\bibinfo {author} {\bibfnamefont {L.~M.}\ \bibnamefont
  {Sandratskii}},\ }\href {\doibase 10.1103/PhysRevB.96.024450} {\bibfield
  {journal} {\bibinfo  {journal} {Phys. Rev. B}\ }\textbf {\bibinfo {volume}
  {96}},\ \bibinfo {pages} {024450} (\bibinfo {year} {2017})}\BibitemShut
  {NoStop}%
\end{thebibliography}

%

\end{document}